# Validating the predictions of mathematical models describing tumor growth and treatment response

Guillermo Lorenzo[1,2,*], David A. Hormuth II[2,3], Chengyue Wu[2,4-7], Graham Pash[2], Anirban Chaudhuri[2], Ernesto A. B. F. Lima[2,8], Lois C. Okereke[2], Reshmi Patel[2], Karen Willcox[2], Thomas E. Yankeelov[2-4,9-11]

[1]Group of Numerical Methods in Engineering, Department of Mathematics, University of A Coruña, Spain
[2]Oden Institute for Computational Engineering and Sciences, The University of Texas at Austin, Austin, TX, USA
[3]Livestrong Cancer Institutes, The University of Texas at Austin, Austin, TX,
[4]Department of Imaging Physics, The University of Texas MD Anderson Cancer Center, Houston, TX, USA
[5]Department of Breast Imaging, The University of Texas MD Anderson Cancer Center, Houston, TX, USA
[6]Department of Biostatistics, The University of Texas MD Anderson Cancer Center, Houston, TX, USA
[7]Institute for Data Science in Oncology, The University of Texas MD Anderson Cancer Center, Houston, TX, USA
[8]Texas Advanced Computing Center, The University of Texas at Austin, Austin, TX, USA
[9]Department of Biomedical Engineering, The University of Texas at Austin, Austin, TX, USA
[10]Department of Diagnostic Medicine, The University of Texas at Austin, Austin, TX, USA
[11]Department of Oncology, The University of Texas at Austin, Austin, TX, USA

*Corresponding author:

Guillermo Lorenzo, PhD
Group of Numerical Methods in Engineering, Department of Mathematics, University of A Coruña
Oden Institute for Computational Engineering and Sciences, The University of Texas at Austin,
Address: E.T.S. de Enxeñeiros de Camiños, Canais e Portos, Universidade da Coruña, Campus de Elviña s/n, 15071 A Coruña, Spain
Email: guillermo.lorenzo@udc.es, guillermo.lorenzo@utexas.edu

# Abstract

Despite advances in methods to interrogate tumor biology, the observational and population-based approach of classical cancer research and clinical oncology does not enable anticipation of tumor outcomes to hasten the discovery of cancer mechanisms and personalize disease management. To address these limitations, individualized cancer forecasts have been shown to predict tumor growth and therapeutic response, inform treatment optimization, and guide experimental efforts. These predictions are obtained *via* computer simulations of mathematical models that are constrained with data from a patient's cancer and experiments. This book chapter addresses the validation of these mathematical models to forecast tumor growth and treatment response. We start with an overview of mathematical modeling frameworks, model selection techniques, and fundamental metrics. We then describe the usual strategies employed to validate cancer forecasts in preclinical and clinical scenarios. Finally, we discuss existing barriers in validating these predictions along with potential strategies to address them.

# Keywords

Predictive oncology, mathematical oncology, mechanistic modeling, tumor forecasting, model selection, uncertainty quantification, model validation, experimental validation, clinical validation, digital twins.

# List of abbreviations

ABM – Agent-Based Model
A/C – Adriamycin® and Cytoxan®
AIC – Akaike Information Criterion
BF – Bayes Factor
BIC – Bayesian Information Criterion
CCC – Concordance Correlation Coefficient
CT – Computerized Tomography
DNA – DeoxyriboNucleic acid
DSC - Dice-Sørensen Coefficient
GSA – Global Sensitivity Analysis
J – Jaccard index
LSA – Local Sensitivity Analysis

MAE – Mean Absolute Error

MAPE – Mean Absolute Percentage Error

MRI – Magnetic Resonance Imaging

NMAE – Normalized Mean Absolute Error

NRMSE – Normalized Root-Mean-Squared Error

ODE – Ordinary Differential Equation

OPAL – Occam Plausibility ALgorithm

PCC – Pearson Correlation Coefficient

pCR – Pathological Complete Response

PDE – Partial Differential Equation

PET – Positron Emission Tomography

PSA – Prostate-Specific Antigen

RMSE – Root-Mean-Squared Error

RNA – RiboNucleic Acid

SA – Sensitivity Analysis

SciML – Scientific Machine Learning

SEER – Surveillance, Epidemiology, and End Results

TCIA – The Cancer Imaging Archive

TNBC – Triple Negative Breast Cancer

# 1. Introduction

Cancer research and clinical decision-making in oncology rely on the measurement of key quantities of interest related to tumor growth and therapeutic response, which are collected according to standardized criteria established in large experimental studies and clinical trials (Cornford et al., 2024; Cowan et al., 2022; Schaff and Mellinghoff, 2023; Waks and Winer, 2019). However, despite the advances in experimental and clinical methods to examine the biophysical mechanisms underlying cancer dynamics (Kazerouni et al., 2020; Penault-Llorca and Radosevic-Robin, 2016; Satar et al., 2021; Willemse et al., 2022), this observational and population-based approach cannot anticipate tumor outcomes. For a disease as heterogeneous as cancer (Almendro et al., 2013; Dagogo-Jack and Shaw, 2018), this limitation hinders the highly coveted personalization of care as well as the discovery of crucial cancer mechanisms for disease monitoring and treatment. These challenging endeavors do not only require characterizing tumor biology at a certain time, but also predicting how tumors would respond to alternative clinical interventions to guide therapeutic decision-making and efficient preclinical studies that advance our knowledge of cancer (Enderling et al., 2019; Gevertz and Kareva, 2024; Kazerouni et al., 2020; Yankeelov et al., 2024).

Tumor forecasting is a computational technology with demonstrated potential to advance the population-based, observational approach of cancer research and clinical oncology towards a personalized, predictive paradigm (Bull and Byrne, 2022; Lorenzo et al., 2023, 2022b; Metzcar et al., 2019; Yankeelov et al., 2024). The last decade has seen rapid growth in the development and application of mathematical models to predict the growth and response of tumors to nearly all possible clinical management strategies, including active surveillance, surgery, radiotherapy, chemotherapy, and immunotherapy (Brady-Nicholls et al., 2020; Hormuth et al., 2021a; Jackson et al., 2015; Lima et al., 2023; Lorenzo et al., 2024a; Schlicke et al., 2021; Wu et al., 2022b). Predictive models generally take as inputs (i) patient-specific initial conditions (e.g., tumor volume, tumor cell density, and tumor cell location and sizes), (ii) model parameters specific to the patient or patient population, and (iii) details of the treatment strategy (Jarrett et al., 2021; Lorenzo et al., 2023, 2022b). Then, computer simulations of the models enable forecasting of tumor growth and response to treatment (Hormuth et al., 2018; Lorenzo et al., 2022b), including key quantities of interest to aid clinicians in decision-making (e.g., changes in tumor burden, tumor extension, and clinical biomarkers). For example, these quantities can be leveraged to predictably identify the best course of clinical management for each individual patient (Brady-Nicholls et al., 2020; Chaudhuri et al., 2023; Jarrett et al., 2020; Lipková et al., 2019). The same forecasting strategy has also been widely used to predict tumor growth and therapeutic response in cell cultures and animal models (Bowers et al., 2022; Gevertz and Kareva, 2024; Lima et al., 2023; Miniere et al., 2024; Strobl et al., 2024), thereby supporting experimental efforts in preclinical cancer research (e.g., experimental study design, investigation of cancer mechanisms, adaptive treatments, and novel drug combinations).

Establishing predictive science in a highly heterogeneous disease such as cancer (Almendro et al., 2013; Dagogo-Jack and Shaw, 2018) demands (i) mathematical precision to reliably represent tumor dynamics under diverse therapeutic conditions and (ii) patient-specific data to realize the model calibration and validate the subsequent predictions. In particular, model validation can be defined as the systematic process of establishing a model's performance and accuracy by comparing its predictions to real-world observations (Brady and Enderling, 2019; Courcelles et al., 2024; Henninger et al., 2010; Jager, 2016; Yankeelov et al., 2024). Hence, model validation for tumor forecasting ensures that the model is reliable and credible in its representation of the disease and treatment response. However, with the heterogeneity in modeling approaches, prediction frameworks, types of cancers, experimental setups, and clinical application scenarios, there is a rapidly developing need for rigorous and standardized approaches for establishing the validity of model predictions. While personalized therapeutic interventions have the potential to improve patient outcomes, reduce off-target toxicity, and deliver patient-centric care (Brady and Enderling, 2019; Chaudhuri et al., 2023; Enderling et al., 2019; Jarrett et al., 2020; Lipková et al., 2019; Strobl et al., 2024; Yankeelov et al., 2024), errors in the predictive models in oncology could directly harm the survival and quality of life of patients. Thus, validation is indispensable to ensure that tumor forecasting models are robust for predicting responses, recommending optimized interventions, and establishing the trustworthiness of the models across disease sites and clinical scenarios as well as among both clinicians and patients.

In this chapter, we present and discuss standard approaches for the validation of tumor forecasting models in preclinical and clinical settings. We first describe common types of mathematical models for tumor forecasting, sensitivity analysis methods to identify critical parameters for model robustness and reliability, and model selection strategies to find the most appropriate model for representing the disease. We also define the common metrics to realize model calibration and assess validation as well as standard methods for uncertainty quantification, which accounts for variability in the model and the observed data to provide a measure of confidence in model predictions. Although not discussed herein, model verification is a necessary previous process that establishes whether the computational implementation of a mathematical model accurately represents its solutions (Henninger et al., 2010; National Research Council et al., 2012; Szabó and Babuška, 2021). Then, we discuss the application of these mathematical and computational approaches to validate predictions of tumor growth and treatment response in both preclinical and clinical scenarios. Finally, the chapter concludes with a discussion of open challenges in validating mathematical models for tumor forecasting and potential strategies to address them.

## 2. Mathematical models of tumor growth and treatment response

### *2.1 Time-resolved models based on ordinary differential equations (ODEs)*

ODE models describe the changes of the tumor burden with respect to time (Lorenzo et al., 2023; Yin et al., 2019). These temporal changes are usually characterized with longitudinal measurements of (for example) tumor volume (Ayala-Hernández et al., 2021; Slavkova et al., 2023; Zahid et al., 2021), cellularity (Barish et al., 2017; Yang et al., 2022), and biomarkers, such as the M-protein for multiple myeloma (Köhn-Luque et al., 2023) and prostate-specific antigen (PSA) for prostate cancer (Brady-Nicholls et al., 2020; Lorenzo et al., 2022a). ODE models describe disease dynamics according to fundamental cancer growth and treatment response mechanisms (Lorenzo et al., 2023; Yin et al., 2019), such as cancer cell proliferation and death (Benzekry et al., 2014; Jarrett et al., 2018b; Johnson et al., 2019) and the cytotoxic effect of therapeutic agents (Ayala-Hernández et al., 2021; Brady-Nicholls et al., 2020; Brüningk et al., 2021; Lima et al., 2022; Zahid et al., 2021). For example, Figure 1A shows a forecast obtained with an ODE model of tumor growth and chemotherapy response (Lorenzo et al., 2023). ODE models can also include the interaction between different cell types, such as tumor and immune cells (Barish et al., 2017; Mirzaei et al., 2022a); tumor cells with different phenotypes, treatment responses, or living in different intratumoral regions (Ayala-Hernández et al., 2021; Brady-Nicholls et al., 2020; Lorenzo et al., 2022a; Slavkova et al., 2023; Strobl et al., 2024; Yang et al., 2022); and the primary tumor and metastases (Bigarré et al., 2023; Schlicke et al., 2021). Beyond the coupled ODEs to describe the dynamics of the cells in the system, additional ODEs can account for temporal changes in key biomolecules (e.g., nutrients and growth factors), vascular density, therapeutic agents, and key tumor biomarkers (Ayala-Hernández et al., 2021; Brady-Nicholls et al., 2020; d'Onofrio et al., 2009; Köhn-Luque et al., 2023; Lima et al., 2022; Lorenzo et al., 2022a; Mirzaei et al., 2022a).

Due to the common availability of longitudinal tumor measurements and the minimal computational demand of ODEs, these models enjoy widespread use and have been validated in multiple experimental and clinical scenarios (see Sections 5 and 6). Their low computational cost is key to comparably assessing alternative treatment regimens and prospectively selecting an optimal personalized therapeutic plan (Ayala-Hernández et al., 2021; Barish et al., 2017; Brady-Nicholls et al., 2020; Brüningk et al., 2021; Gevertz and Kareva, 2024; Lima et al., 2022; Strobl et al., 2024). Additionally, these low-cost models facilitate the propagation of data and model uncertainties in the tumor forecasts to support clinical decision-making, which heavily relies on probabilistic risks and survival odds (Brady-Nicholls et al., 2020; Chaudhuri et al., 2023; Zahid et al., 2021). However, the main disadvantage of ODE models is their inability to describe spatially-resolved changes in tumor morphology and the mechanisms governing them.

*2.2 Spatiotemporally-resolved models based on partial differential equations (PDEs)*

PDE models represent the growth and treatment response of tumors over time and across a spatial region of interest (Lorenzo et al., 2023), such as a Petri dish (Miniere et al., 2024), a hexahedral domain enclosing a tumor spheroid or an *in vivo* tumor (Bowers et al., 2022; Lorenzo et al., 2024b; Roose et al., 2003; Wong et al., 2016), and patient-specific organ geometries (Chen et al., 2012; Hormuth et al., 2021a; Jackson et al., 2015; Lipková et al., 2019; Lorenzo et al., 2024a; Wu et al., 2022b). The tumor can be modeled *via* tumor cell density and volumetric fractions (Chen et al., 2012; Hormuth et al., 2021a; Kremheller et al., 2019; Lipková et al., 2019; Lorenzo et al., 2024a; Urcun et al., 2021; Wu et al., 2022b), tumor phase fields (Agosti et al., 2018; Fritz et al., 2021; Lima et al., 2017; Lorenzo et al., 2019a; Wise et al., 2008), and deformable tissue domains (Roose et al., 2003; Stylianopoulos et al., 2013; Vavourakis et al., 2017). These models include mobility mechanisms usually consisting of diffusion and advection operators, which are either explicitly designed for the model or emerge from the minimization of energy functionals (e.g., in phase-field models (Fritz et al., 2021; Lima et al., 2017; Wise et al., 2008)). PDE models further include reaction terms for tumor cell proliferation, cytotoxic action of treatments, and interactions between cancerous and non-cancerous cell species that are analogous to those in ODE models (Agosti et al., 2018; Fritz et al., 2021; Hormuth et al., 2021a; Lima et al., 2017; Mirzaei et al., 2022b; Swanson et al., 2011; Wise et al., 2008; Wu et al., 2022b). Likewise, PDE models can be posed in a multicellular multicompartmental framework (Fritz et al., 2021; Miniere et al., 2024; Mirzaei et al., 2022b; Swanson et al., 2011; Wise et al., 2008), and they can also feature PDEs to describe the spatiotemporal dynamics of vascular density (Hormuth et al., 2021c, 2019a; Kremheller et al., 2019; Vilanova et al., 2017; Xu et al., 2020), therapeutic agents (Hormuth et al., 2021a; Lipková et al., 2019; Lorenzo et al., 2024b; Vavourakis et al., 2018; Wu et al., 2022b), and other key tumor-related variables, such as nutrients, growth factors, and biomarkers (Lorenzo et al., 2019a; Mirzaei et al., 2022b; Swanson et al., 2011; Vavourakis et al., 2017; Wise et al., 2008; Xu et al., 2020). Additionally, the mechanical deformation of the tumor and neighboring tissue can be coupled to tumor dynamics to account for the growth-arresting effects of solid stress and interstitial pressure (Blanco et al., 2023; Bowers et al., 2022; Hormuth et al., 2021c; Lorenzo et al., 2019a; Roose et al., 2003; Stylianopoulos et al., 2013; Urcun et al., 2021; Vavourakis et al., 2017; Wong et al., 2016; Wu et al., 2022b). For example, Figure 1B shows a spatiotemporal tumor forecast obtained with a PDE model describing the growth and the response of breast cancer to neoadjuvant chemotherapy (Patel et al., 2024; Wu et al., 2022b).

The spatiotemporal data to personalize PDE models consists of longitudinal imaging measurements (Corwin et al., 2013; Hormuth et al., 2018; Lorenzo et al., 2022b). However, time-resolved tumor and biomarker data can also be leveraged to calibrate and validate these models (Colli et al., 2021; Lorenzo et al., 2019a; Roose et al., 2003; Swanson et al., 2011; Urcun et al., 2021; Vavourakis et al., 2017; Wu et al., 2022b; Xu et al., 2020), since spatial integration of their main variables yields those modeled with ODEs (e.g., tumor volume, total cellularity, and biomarkers). While PDE models address the lack of a spatial component in ODE models, their spatiotemporal formulation requires more computationally demanding numerical methods, such as finite differences, finite elements, and isogeometric analysis (Hormuth et al., 2018; Lorenzo et al., 2022b). This issue has not impeded the validation of PDE models in several experimental and clinical scenarios (see Sections 5 and 6), nor their use for treatment optimization (Colli et al., 2021; Corwin et al., 2013; Jarrett et al., 2020; Lipková et al., 2019; Wu et al., 2022a) and uncertainty quantification (Lima et al., 2017; Lipková et al., 2019; Mascheroni et al., 2021). However, the high computational demand per model simulation can dramatically hinder these latter two applications of PDE models, which are pivotal for clinical decision-making.

## 2.3 Agent-based models (ABMs)

ABMs describe the spatiotemporal evolution of a tumor and its microenvironment by representing the main cell types in the system (e.g., healthy, tumor, immune, and endothelial cells) as discrete agents that reside and interact in a small region of tissue or *in vitro* setting (Metzcar et al., 2019; Van Liedekerke et al., 2015; West et al., 2023). These agents can be geometrically defined upon the discretization of the domain in lattice-based models (Bravo et al., 2020; Burbanks et al., 2023; Sosa-Marrero et al., 2021; Strobl et al., 2022; Szabó and Merks, 2013; Vega et al., 2020), or as entities with prescribed shape that freely roam in space in lattice-free models (Ghaffarizadeh et al., 2018; Palmieri et al., 2015; Pérez-Velázquez and Rejniak, 2020; Phillips et al., 2023a; Rocha et al., 2018). The agents' behavior is governed by user-defined rules that combine deterministic and probabilistic formulations to dictate, for example, how cells proliferate, move, and respond to treatment. The dynamics of key substances (e.g., nutrients, growth factors, and drugs) and mechanical equilibrium are usually described with PDEs (Bravo et al., 2020; Burbanks et al., 2023; Ghaffarizadeh et al., 2018; Palmieri et al., 2015; Pérez-Velázquez and Rejniak, 2020; Rocha et al., 2018; Sosa-Marrero et al., 2021; Vega et al., 2020). Subcellular processes can be represented with ODEs for each agent, such as protein pathways regulating cell behavior, therapeutic damage, and chemoresistance (Pérez-Velázquez and Rejniak, 2020; Rocha et al., 2018; Vega et al., 2020). For instance, Figure 1C illustrates the output of a lattice-free model in which cancer cells progressively die due to the scarce availability of nutrients (Lima et al., 2021).

ABMs provide a richer description of spatial tumor heterogeneity than PDE models, while also accounting for the natural stochasticity of tumor biology (Metzcar et al., 2019; Van Liedekerke et al., 2015; West et al., 2023). However, this level of granularity requires lower spatial and/or temporal scales than PDEs. Furthermore, agent-based data to inform these models is usually scarce since it requires large experimental efforts or non-standard histological processing. Thus, although imaging data (Paczkowski et al., 2021; Phillips et al., 2023a) and histopathological data (Macklin et al., 2012; Sosa-Marrero et al., 2021) have been used to inform these models, spatially-averaged metrics derived from these data types are more commonly employed for this purpose (e.g., tumor volume, total cellularity, microscopy confluence, and biomarkers; see Section 5).

## 3. Tools for model selection

### 3.1 Sensitivity analysis

Sensitivity analysis (SA) is a crucial step in model development that aims at quantifying how changes in model input factors (e.g., parameters) impact model outputs of interest (Craig et al., 2023; Lorenzo et al., 2022b; Qian and Mahdi, 2020; Saltelli et al., 2008, 1999; Sobol, 2001). Hence, SA identifies the most relevant inputs of the model, which require accurate calibration from data and indicate which model mechanisms (related to those inputs) are driving the overall dynamics of the outputs of interest. Thus, SA is a fundamental tool for model development, validation, and selection (Craig et al., 2023; Lorenzo et al., 2022b; Oden et al., 2017). SA methods can be broadly classified into two categories: local sensitivity analysis (LSA) and global sensitivity analysis (GSA). LSA assesses how perturbations of each single input parameter (i.e., first-order effects) affect model outputs (Craig et al., 2023; Qian and Mahdi, 2020). The most common LSA methods are derivative-based, whereby the sensitivity index $S_i$ of the $i$-th parameter, $p_i \in \theta = \{p_1, p_2, \ldots, p_n\}$, to the model output $Y = \phi(x, t, \theta)$ is calculated as

$$S_i = \gamma_i \frac{\partial \phi}{\partial p_i} \approx \gamma_i \frac{\phi(x, t, p_1, p_2, \ldots, p_i + \Delta p_i, \ldots, p_n) - \phi(x, t, p_1, p_2, \ldots, p_i, \ldots, p_n)}{\Delta p_i}, \qquad [1]$$

where $\Delta p_i$ is the perturbation of the $i$-th parameter and $\gamma_i$ is a scaling factor, which can account for reference values or variances of input parameters and model outputs (Craig et al., 2023; Qian and Mahdi, 2020). The major advantage of LSA is its simplicity, but it neglects the interaction between input parameters (i.e., high-order effects). Conversely, GSA quantifies the impact of variations of all parameters on the model outputs (Craig et al., 2023; Saltelli et al., 2008, 1999; Sobol, 2001). Variance-based methods are the most common GSA technique, and they rely on a decomposition of the variance of a model output ($V(Y)$) as a sum of the contribution to this variance from each parameter and all possible parameter combinations of parameters in $\theta$. Then, the first-order and total sensitivity indices ($S_i$ and $T_i$, respectively) for each parameter $p_i$, can be calculated as

$$S_i = \frac{V_i}{V(Y)} = \frac{V(E(Y|p_i))}{V(Y)}, \quad T_i = 1 - \frac{V(E(Y|p_{-i}))}{V(Y)}. \qquad [2]$$

In Eq. [2], $E(Y|p_i)$ denotes the conditional expectation of output $Y$ given $p_i$, and $p_{-i}$ denotes all parameter combinations without $p_i$ (Saltelli et al., 2010, 2008). Since GSA can entail a high computational cost, several efficient estimator algorithms to calculate $S_i$ and $T_i$ have been proposed (Piano et al., 2021; Saltelli et al., 2010). Of note, both LSA and GSA have been successfully and extensively applied in the development and validation of tumor forecasting models (Brady-Nicholls et al., 2020; Brüningk et al., 2021; Jenner et al., 2021; Lima et al., 2023, 2021; Lorenzo et al., 2024b; Poleszczuk et al., 2015; Urcun et al., 2021).

## *3.2 Information criteria*

Information criteria are the cornerstone of model selection methods. These criteria systematically evaluate competing models in a candidate set and enable the identification of a model that optimally balances goodness-of-fit and formulation simplicity (Burnham and Anderson, 2002, 2004; Kass and Raftery, 1995; Konishi and Kitagawa, 2008; Resende et al., 2022). The most widely used in mathematical oncology are the Akaike Information Criterion (AIC) and the Bayesian Information Criterion (BIC).

The AIC is rooted in information theory (Burnham and Anderson, 2004; Konishi and Kitagawa, 2008), and it is defined as

$$AIC = -2\log \pi(d|\hat{\theta}) + 2k, \qquad [3]$$

where $\pi(d|\hat{\theta})$ is the probability of observing data $d$ given the maximum likelihood estimate $\hat{\theta}$ of the model parameters, while $k$ is the number of parameters in the model and acts as a penalty term to limit model complexity. The AIC minimizes the Kullback–Leibler divergence, which quantifies the discrepancy between the fitted model and the true underlying data-generating process (Burnham and Anderson, 2002). Consequently, the AIC selects models that are efficient predictors, particularly in small sample sizes. However, the AIC is not consistent, since it may fail to select the true model, especially in large-sample scenarios (Burnham and Anderson, 2002, 2004).

The BIC relies on statistical inference (Burnham and Anderson, 2004; Kass and Raftery, 1995), and it is defined as

$$BIC = -2\log \pi(d|\hat{\theta}) + k\log(n), \qquad [4]$$

where $n$ is the number of observed data. The inclusion of $\log(n)$ in the penalty term results in a stronger penalization for model complexity compared to the AIC. Hence, the BIC favors simpler models as the sample size increases, thereby achieving consistency in selecting the correct model when $n \to \infty$ (Burnham and Anderson, 2004). This property ensures that, given enough data, the BIC will select the true model if it exists in the candidate set (Burnham and Anderson, 2002, 2004; Kass and Raftery, 1995; Resende et al., 2022).

While both AIC and BIC are widely applicable, their performance depends on data size and the study goals. In small-sample scenarios, AIC may perform better by selecting models with stronger predictive power. As the amount of observed data grows, BIC's stronger penalization ensures the selection of simpler, more parsimonious models (Burnham and Anderson, 2002; Konishi and Kitagawa, 2008; Resende et al., 2022). Thus, the decision between these criteria should be guided by whether the analysis prioritizes predictive accuracy (AIC) or parsimonious inference of the true data-generating model (BIC).

### *3.3 Ensemble modeling*

Instead of relying on a single best model, ensemble modeling consists of combining predictions from multiple candidate models (Hormuth et al., 2021b; Parker, 2013). This approach can potentially address model uncertainty while improving accuracy and precision. A common strategy for ensemble modeling is model averaging (Hormuth et al., 2021b), whereby model outputs are weighted based on their performance using information criteria (e.g., AIC or BIC). These model weights are defined as:

$$w_i = \frac{\exp(-\Delta_i/2)}{\sum_{r=1}^{m}\exp(-\Delta_r/2)}, \qquad [5]$$

where $m$ is the number of models and $\Delta_i = IC_i - IC_{min}$ is the difference between the chosen information criterion score $IC$ for each model with respect to the best one (i.e., with minimum $IC$, $IC_{min}$). These weights are normalized such that $\sum_{i=1}^{m} w_i = 1$, thus providing a probabilistic interpretation of the information criterion. Then, the ensemble prediction can be constructed as

$$\phi_{ens}(x,t,\theta) = \sum_{i=1}^{m} w_i \phi_i(x,t,\theta), \qquad [6]$$

where, for each spatial point $x$, time $t$, and parameters $\theta$, $\phi_{ens}(x,t,\theta)$ denotes the ensemble-averaged spatiotemporal tumor map while $\phi_i(x,t,\theta)$ is the prediction from each model (Hormuth et al., 2021b). Similar expressions can be derived for time-resolved ODE models as well as PDE and ABM outputs.

## 4. Methods for the validation of cancer model forecasts

### *4.1 Global and local model quantities for validation*

Several quantities of interest derived from the model constituents (i.e., variables, parameters) can be leveraged for the validation of mechanistic models of cancer. The choice depends on the available data, since the quantities of interest need to match the biophysical nature of the measurements. Additionally, the quantities of interest can be global or local. Global quantities describe the tumor burden as a whole entity and do not carry spatial resolution; for example, tumor volume (Ayala-Hernández et al., 2021; Brüningk et al., 2021; Hormuth et al., 2021a; Zahid et al., 2021), tumor cellularity or cell counts (Burbanks et al., 2023; Chaudhuri et al., 2023; Lorenzo et al., 2024a; Miniere et al., 2024; Mirzaei et al., 2022a; Wu et al., 2022b), vascular density (Phillips et al., 2023a), global proliferation activity (Ayala-Hernández et al., 2021; Bosque et al., 2023; Lorenzo et al., 2024a), and tumor-specific biomarkers, such as PSA (Brady-Nicholls et al., 2020; Lorenzo et al., 2022a) and the M-protein (Köhn-Luque et al., 2023). Local quantities of interest are defined pointwise over the spatial region occupied by the tumor, such as tumor cell density (Bowers et al., 2022; Hormuth et al., 2021a; Lorenzo et al., 2024a; Wong et al., 2016; Wu et al., 2022b), vascularity or vascular density map (Hormuth et al., 2019a; Phillips et al., 2023a), tumor phase field (Agosti et al., 2018; Lima et al., 2017), and mechanical deformation (Bowers et al., 2022; Stylianopoulos et al., 2013). Thus, while ODE model validation relies on global quantities, PDE models and ABMs can use both global and local quantities for validation. However, local metrics provide a higher level of validation for spatially-resolved models since they assess whether the model prediction matches the data at each point in space.

## *4.2 Deterministic model validation*

Deterministic validation assesses whether model predictions obtained with a single set of parameters exhibit an acceptable agreement with measurements different to those used for parameter calibration, which consists of an inverse problem usually formulated *via* nonlinear least-squares methods and PDE/ODE-constrained optimization methods (Henninger et al., 2010; Hormuth et al., 2018; Lorenzo et al., 2022b; Nocedal and Wright, 1999; Szabó and Babuška, 2021). The error metrics used for model validation are usually included in the objective functional that is minimized during parameter calibration. In mathematical oncology, the most common error metrics are the root-mean-squared error (RMSE), the mean absolute error (MAE), and the mean absolute percentage error (MAPE) (Ayala-Hernández et al., 2021; Brüningk et al., 2021; Ghaffari Laleh et al., 2022; Lima et al., 2017, 2022; Lorenzo et al., 2024a, 2022a; Martens et al., 2022; Wong et al., 2016). These error metrics are defined as

$$RMSE = \sqrt{\frac{1}{n}\sum_{i=1}^{n}\bigl(Y(t_i) - X(t_i)\bigr)^2}, \qquad [7]$$

$$MAE = \frac{1}{n}\sum_{i=1}^{n}|Y(t_i) - X(t_i)|, \qquad [8]$$

$$MAPE = \frac{100}{n}\sum_{i=1}^{n}\left|\frac{Y(t_i) - X(t_i)}{X(t_i)}\right|, \qquad [9]$$

where $Y(t_i)$ and $X(t_i)$ denote the model prediction and the measurement of a certain quantity of interest at time $t_i$ (see Fig. 2). For spatially-resolved models, $Y(t_i)$ and $X(t_i)$ are calculated over the spatial region occupied by the tumor through summation (ABMs) or spatial integration (PDE model and ABMs). While MAPE is a relative metric defined in the range [0, 100], RMSE and MAE are absolute metrics that need to be assessed in the context of each specific forecasting scenario. Normalization of RMSE and MAE with the mean, median, range, or maximum value of the measurements results in corresponding relative metrics (NRMSE and NMAE) (Flores-Torres et al., 2023; Yang et al., 2022; Zahid et al., 2021).

Validation of spatially-resolved models usually includes metrics examining the model-data agreement in tumor shape. The most common ones are the Dice-Sørensen coefficient ($DSC$) (Dice, 1945; Hormuth et al., 2021a; Lorenzo et al., 2024a; Sorensen, 1948; Wong et al., 2016; Zhang et al., 2025) and the Jaccard index ($J$) (Agosti et al., 2018; Jaccard, 1912; Swan et al., 2018), which are defined as

$$DSC = \frac{2|\Omega_Y(t_i) \cap \Omega_X(t_i)|}{|\Omega_Y(t_i)| + |\Omega_X(t_i)|}, \qquad [10]$$

$$J = \frac{|\Omega_Y(t_i) \cap \Omega_X(t_i)|}{|\Omega_Y(t_i) \cup \Omega_X(t_i)|}, \qquad [11]$$

where $\Omega_Y(t_i)$ and $\Omega_X(t_i)$ denote the spatial region defined by a quantity of interest at time $t_i$ as predicted by the model and measured from data, respectively. Based on their definitions in Eqs. [10] and [11], $DSC$ and $J$ are connected through the formula $J = DSC/(2 - DSC)$.

Correlation coefficients are also leveraged for validation in mathematical oncology. The most common ones are the Pearson correlation coefficient ($PCC$), which assesses if the values of a quantity of interest from data and model align over a straight line, and the concordance correlation coefficient ($CCC$), which specifically measures their alignment along the line of unity (Lin, 1989). These two correlation coefficients are defined as

$$PCC = \frac{CV(X,Y)}{\sigma(X)\sigma(Y)} \qquad [12]$$

$$CCC = \frac{2\,CV(X,Y)}{V(X) + V(Y) + (\bar{X} - \bar{Y})^2}, \qquad [13]$$

where $X$ and $Y$ denote the values obtained from measurements and the model, respectively. In Eqs. [12] and [13], $CV(X,Y)$ is the covariance of $X$ and $Y$, while $\sigma(X)$ denotes the standard deviation of $X$. The $PCC$ and $CCC$ can be used globally, to assess model-data agreement across patients in a cohort using global quantities of interest, and locally, to assess spatial model-data agreement for an individual tumor (Hormuth et al., 2021a; Jarrett et al., 2018a; Lorenzo et al., 2024a). Another commonly used correlation metric is the coefficient of determination $R^2$ (Brady-Nicholls et al., 2020; Stylianopoulos et al., 2013; Yang et al., 2022), which is defined as

$$R^2 = 1 - \frac{\sum_{i=1}^{n}\big(Y(t_i) - X(t_i)\big)^2}{\sum_{i=1}^{n}(X(t_i) - \bar{X})^2}, \qquad [14]$$

where $\bar{X}$ denotes the mean of the measurements $X(t_i)$.

### *4.3 Bayesian model validation and uncertainty quantification*

The continuous assessment of uncertainty in tumor forecasting as patient data is collected during monitoring and treatment as well as the validation of cancer models under uncertainty are essential to ensuring their reliability and building trust in their clinical application (Chaudhuri et al., 2023; Hawkins-Daarud et al., 2013; Lima et al., 2017; Lipková et al., 2019; Lorenzo et al., 2022b; Oden et al., 2016; Wu et al., 2022c). Bayesian model calibration provides a framework to account for the different sources of uncertainty that impact estimates of the model parameters based on noisy observed data as well as correcting for any inadequacy of the model itself (Biegler et al., 2011; Ghattas and Willcox, 2021; Kaipio and Somersalo, 2006; Kennedy and O'Hagan, 2001; Oden et al., 2017; Smith, 2024; Stuart, 2010). The Bayesian formulation combines prior knowledge with observed data to update a probabilistic distribution of the model parameters. The solution to the inverse problem in Bayesian calibration is the posterior distribution $\pi_{\text{post}}(\theta|d)$, which describes the distribution of the model parameters, $\theta$, conditioned on the observed data, $d$. The posterior distribution can be defined through application of Bayes rule as

$$\pi_{\text{post}}(\theta|d) = \frac{\pi_{\text{like}}(d|\theta)\pi_{\text{pr}}(\theta)}{\int \pi_{\text{like}}(d|\theta)\pi_{\text{pr}}(\theta)d\theta}. \qquad [15]$$

The prior distribution $\pi_{\text{pr}}(\theta)$ quantifies the knowledge of the parameters before calibration, such as from the population level or previous model calibrations (see Fig. 3). The likelihood $\pi_{\text{like}}(d|\theta)$ is specific to the choice of noise model and measures how well model predictions given parameters match the observational data. The denominator is the evidence and normalizes the resultant posterior distribution. Then, the uncertainty in the parameters is propagated through the model by sampling realizations of the model parameters from the posterior and solving the forward model (see Fig. 3). This procedure yields the posterior predictive distribution of the model outputs (i.e., variables, quantities of interest), which can be used for model validation and decision-making under uncertainty (Chaudhuri et al., 2023; Lima et al., 2017, 2023, 2022; Lipková et al., 2019; Mascheroni et al., 2021; Pasetto et al., 2021; Phillips et al., 2023a).

To assess whether a model with quantified uncertainty is valid for a patient, error and correlation metrics (see Section 4.2) can be applied by using estimates of a specific statistic of the posterior predictive distribution of desired model outputs, such as the mean or the standard deviation (Ezhov et al., 2021; Lima et al., 2023, 2022; Mascheroni et al., 2021), and comparing against the statistics of the observed data. Another validation strategy is by using the Bayes factor in the context of Bayesian hypothesis testing (Berger and Delampady, 1987; Ling and Mahadevan, 2013; O'Hagan, 1995). Denoting $H_0$ and $H_1$ as the null and alternative hypotheses of a model being acceptable or not, the Bayes factor $BF$ is then defined as

$$BF = \frac{P[d \mid H_1]}{P[d \mid H_0]}. \qquad [16]$$

Hence, if $BF \gg 1$, the data provides strong evidence for the validity of the model, whereas, if $BF \ll 1$, the data strongly supports that the model is not valid. Nevertheless, the choice of alternate hypothesis and prior beliefs on the hypothesis play a significant role in the Bayes factor estimates, which can make it difficult to analyze in a practical setting for varying validation data.

To validate a model under uncertainty, it is also possible to use the distance between the posterior predictive distribution of a model output and the distribution associated with the observed validation data. There are several metrics for measuring the distance between probability density functions such as Kullback–Leibler divergence ($D_{\text{KL}}$), the Jensen-Shannon distance ($D_{\text{JS}}$), and the Wasserstein distance ($W$) (Endres and Schindelin, 2003; Osterreicher and Vajda, 2003; Panaretos and Zemel, 2019), which are respectively defined as

$$D_{\text{KL}}(P||Q) = \int p(x) \log \frac{p(x)}{q(x)} dx, \qquad [17]$$

$$D_{\text{JS}}(P,Q) = \sqrt{\frac{1}{2}\big(D_{\text{KL}}(P||M) + D_{\text{KL}}(Q||M)\big)}, \quad M = \frac{1}{2}(P+Q) \qquad [18]$$

$$W_r(P,Q) = \left(\inf_{\gamma \in \Gamma(P,Q)} \int d(x,y)^r d\gamma(x,y)\right)^{1/r}, \qquad [19]$$

where $P$ denotes the posterior distribution of a quantity of interest from the model and $Q$ is the distribution of the validation data in the context of cancer model validation. In Eq. [17], the probability densities of $P$ and $Q$ are represented by $p$ and $q$, respectively. Additionally, in Eq. [19], $r$ is the order of the Wasserstein distance, $\Gamma(P,Q)$ is the set of all joint distributions $\gamma(x,y)$ such that marginal of $\gamma$ over $x$ is $P$ and marginal of $\gamma$ over $y$ is $Q$, and $d(x,y)$ is the distance between $x$ and $y$. The interested reader is referred to (Endres and Schindelin, 2003; Osterreicher and Vajda, 2003; Panaretos and Zemel, 2019) for further detail on these metrics. Of note, $D_{JS}$ and $W_p$ are symmetric metrics that are more suitable for continuous validation of tumor forecasts and their dynamic updates with incoming data during cancer monitoring and treatment.

## 5. Model validation in the preclinical setting

### 5.1 Experimental validation of tumor forecasting models

The goal of the preclinical validation of a mathematical model of cancer is to ensure that it accurately captures fundamental physical and biological processes of interest in its formulation. This can be accomplished by using data collected in experiments that isolate cancer dynamics *in vitro* (e.g., cell cultures, tumor spheroids, and microfluidic devices) (Bowers et al., 2022; Johnson et al., 2019; Lorenzo et al., 2024b; Miniere et al., 2024; Paczkowski et al., 2021; Phillips et al., 2023a; Roose et al., 2003; Urcun et al., 2021; Yang et al., 2022) or in more realistic *in vivo* settings (e.g., animal models) (Barish et al., 2017; Burbanks et al., 2023; Cerasuolo et al., 2020; Hormuth et al., 2019a; Lima et al., 2017, 2023, 2022; Mirzaei et al., 2022b, 2022a; Segura-Collar et al., 2022; Slavkova et al., 2023; Stylianopoulos et al., 2013). Although the preclinical setting may not fully capture the complex interactions of clinical disease, it provides a more controlled environment for quantitative validation of mathematical models of tumor growth and response. For example, preclinical studies allow more frequent observation, untreated tumor growth, and precise control over biological variables, thereby facilitating rigorous testing and refinement of these models. In doing so, modelers can build confidence and trust in their models and methods before moving to *in vivo* clinical settings where confounding effects abound (e.g., data sparsity, tumor-microenvironment interactions, heterogeneity within and across patients). For example, Figure 4 illustrates the process of validating the forecasts and optimal treatment plans obtained with an ODE model of therapeutic response of breast cancer using data from mice (Lima et al., 2022).

### 5.2. Data for model validation in the preclinical setting

The choice of *in vivo* versus *in vitro* preclinical studies is largely dictated by the type of model and quantities of interest under investigation (see Section 2) (Kazerouni et al., 2020; Sailer et al., 2023). For example, longitudinal *in vivo* studies are often suitable for quantifying tissue scale or whole-tumor properties in animals with an intact vasculature network and with (or without) a functional immune system (Barish et al., 2017; Burbanks et al., 2023; Cerasuolo et al., 2020; Hormuth et al., 2019a; Lima et al., 2017, 2023, 2022; Mirzaei et al., 2022b, 2022a; Segura-Collar et al., 2022; Slavkova et al., 2023; Stylianopoulos et al., 2013). Common data types available at this scale include longitudinal medical imaging data (e.g., magnetic resonance imaging; MRI, positron emission tomography; PET, and computed tomography; CT), tumor caliper measurements (e.g., tumor volume measurements), blood draws (e.g., circulation biomarkers and drug concentration), as well as post-mortem histology and transcriptomics. Of these methods, only medical imaging data provides non-invasive, longitudinal, 3D intra-tumor measurements; for example, cellularity, perfusion, oxygenation, metabolism, and tissue structure (Hidrovo et al., 2017; Hormuth et al., 2019b, 2019a; Lima et al., 2017, 2023; Slavkova et al., 2023). Caliper measurements and blood draws may also provide longitudinal assessment of whole tumor or systemic properties (Barish et al., 2017; Lima et al., 2023, 2022). Histological samples can provide detailed 1D, 2D, or limited 3D spatial information about tumor structure and composition, while also enabling gene and protein expression analyses (Burbanks et al., 2023; Cerasuolo et al., 2020; Mirzaei et al., 2022a, 2022b; Segura-Collar et al., 2022; Stylianopoulos et al., 2013). However, these histological samples are usually endpoint data collected after tissue harvesting post-mortem. Alternatively, *in vitro* studies are best suited for observing and quantifying cell counts, cell-scale responses and interactions, signaling pathways, as well as gene and protein expression (Bowers et al., 2022; Johnson et al., 2019; Lorenzo et al., 2024b; Miniere et al., 2024; Paczkowski et al., 2021; Phillips et al., 2023a; Roose et al., 2003; Urcun et al., 2021; Yang et al., 2022). Common data types available at this scale include longitudinal time-resolved microscopy, assays assessing gene expression and protein content (e.g., polymerase chain reaction, RNA sequencing, and flow cytometry), and microfluidic platforms (Kazerouni et al., 2020; Sailer et al., 2023). However, with the exception of longitudinal time-resolved microscopy, all of these approaches are generally limited to endpoint data or require take-down studies to observe the dynamics of, for instance, gene expression in response to therapy.

As with any experimental measure, both *in vivo* and *in vitro* studies in cancer research have several limitations (Kazerouni et al., 2020; Sailer et al., 2023). Firstly, they contain uncertainty in the quantification of the biological property due to instrument precision, indirect measurement methods, and variations in experimental procedure. As such, rigorous efforts should be made to select measurement techniques that have established repeatability and reproducibility studies to understand the uncertainty within the data underlying any model. Secondly, a unique limitation of most preclinical studies is that cancer cells used to initialize *in vivo* or *in vitro* studies are clones of each other, which dramatically limits the amount of intratumor heterogeneity observed compared to the clinical disease. Finally, there is currently a paucity of publicly available preclinical data required for validating mathematical models of tumor growth and response. Thus, most preclinical model validation studies may require the acquisition of new data as there are limited publicly available longitudinal datasets in resources such as The Cancer Imaging Archive (TCIA) or the Cancer Research Data Commons (Clark et al., 2013; Fedorov et al., 2021; Wang et al., 2024).

## 5.3. Examples of model validation studies in the preclinical setting

### 5.3.1 In vitro validation of an alternative model for low-density tumor growth

Model selection methods (see Section 3) may be used in validation studies to determine more plausible cancer models. For example, the study in (Johnson et al., 2019) investigated alternative models of tumor growth at low density (e.g., in early stages of the disease and post-treatment growth) in comparison to the conventional exponential growth model by leveraging an ODE modeling framework informed with *in vitro* data. The authors built upon prior observations that revealed a discrepancy between the exponential model and tumor growth at low cell densities. Drawing inspiration from ecology, they hypothesized that the Allee effect could be a more representative modeling approach (Courchamp et al., 2008). The Allee effect arises from cooperative growth and manifests as a scaling of growth rate with the size of the tumor cell population when this is small. To experimentally test this hypothesis, BT-474 breast cancer cells were seeded with 1 to 20 cells per well of a 96-well plate and time-lapse microscopy images were collected every 4 hours over a total of 328 hours. Statistical tests on the resulting data confirmed significantly lower growth rates in the time-resolved measurements of low initial cell count colonies. Leveraging this experimental evidence, the authors then used Bayesian model selection to compare the traditional exponential model against several extensions of this model including the Allee effect as modifications to the underlying tumor cell birth and death processes (i.e., thus making each or both of them dependent on the population size). To this end, they employed Bayesian calibration (see Section 4.3) and the BIC for model selection (see Section 3.2). The results showed that the model that more accurately represented the experimental data incorporated the Allee effect as an increase of the tumor cell birth rate with population size in the small population regimen. Thus, the study in (Johnson et al., 2019) integrated many aspects of model validation in the preclinical setting to deliver insight into the fundamental nature of tumor growth.

### 5.3.2 Validation of PDE models of tumor radiotherapy response using in vivo murine data

*In vivo* model validation represents a critical step in ensuring the robustness of predictive mathematical frameworks for tumor growth and treatment response. To this end, the work in (Lima et al., 2017) conducted a model selection and validation study to identify the best formulation describing brain tumor growth and response to radiotherapy according to longitudinal MRI data collected from rats. The modeling framework is based on the Occam Plausibility Algorithm (OPAL) (Lorenzo et al., 2022b; Oden et al., 2017), which leverages Bayesian principles to calibrate, validate, and select models against experimental observations (see Section 4.3). The authors proposed a family of thirty-nine models, which resulted from combining three models of radiotherapeutic effects and thirteen tumor growth models. Using the MRI measurements of brain tumor growth and radiotherapy response in rats, they calibrated their models to predict the effects of a 20 Gy or 40 Gy radiation dose, accounting for uncertainties in the data and model parameters. A significant aspect of this study is the inclusion of radiotherapy models that capture delayed tumor cell death post-treatment. Specifically, unlike classical linear-quadratic models assuming instantaneous effects of radiation-induced tumor cell death, this work incorporates memory and partial-memory models to better represent the radiotherapeutic response observed *in vivo*. Among the evaluated models, those integrating memory effects achieved the highest validation accuracy, demonstrating their superior predictive capacity for therapy-induced tumor reduction. Thus, the study in (Lima et al., 2017) exemplifies how rigorous *in vivo* validation, combined with Bayesian calibration techniques accounting for model and data uncertainties, enables the refinement of mathematical models to capture the inherent complexity of biological processes underlying cancer growth and treatment response.

## 6. Model validation in the clinical setting

### *6.1 Clinical validation of tumor forecasting models*

The validation of mathematical models of cancer in clinical settings aims to ensure that patient-specific forecasts of tumor growth and treatment response have a consistent agreement with the corresponding clinical observations in patients, so that these predictions can be used to address clinical questions reliably (Enderling et al., 2019; Wu et al., 2022c; Yankeelov et al., 2024). Hence, depending on the final purpose of the model and its level of readiness, validation studies can be performed with different specific endpoints. The development of clinically-oriented tumor forecasting models usually starts by utilizing pre-clinical data and/or small clinical datasets with well curated conditions for proof-of-principle examinations (Agosti et al., 2018; Hormuth et al., 2019a; Jarrett et al., 2018a; Lorenzo et al., 2019a; Rockne et al., 2010; Urcun et al., 2023). Then, validation of tumor forecasting accuracy is commonly established using increasingly larger retrospective cohorts. These validation studies compare the model outputs to the "ground truth" of clinically relevant metrics and test whether there are no statistically significant differences between them. For example, a model that seeks to predict tumor response to certain therapy can be validated by comparing to the clinical endpoints of treatment outcomes, such as pathological response status, overall survival rate, progress-free survival rate, or time-to-progression (Brady-Nicholls et al., 2020; Chaudhuri et al., 2023; Chen et al., 2012; Hormuth et al., 2021a; Lipková et al., 2019; Lorenzo et al., 2024a, 2022a; Wong et al., 2016; Wu et al., 2022b; Zhang et al., 2025). Once tumor forecasting accuracy is validated within the discovery cohort on which the model is built, external validation will be needed to assess the model's generalizability to other cohorts. Hence, external validation aims to prove that the accuracy of the model can be decently preserved on datasets collected from different institutions, with different data collection and clinical management protocols or even within different patient populations (Bigarré et al., 2023; Bosque et al., 2023; Chen et al., 2012; Jarrett et al., 2020; Lorenzo et al., 2022a, 2019b; Wong et al., 2016; Wu et al., 2022b). Uncertainty quantification during internal and external validation seeks to establish the reliability of the tumor forecasts, enabling the calculation of probabilistic and risk metrics that are central to clinical decision-making (e.g., malignant progression and post-treatment tumor recurrence) (Brady-Nicholls et al., 2020; Chaudhuri et al., 2023; Lipková et al., 2019; Lorenzo et al., 2022a). At this stage, model validation studies further investigate the presence of sources of biases causing any remarkable differences across different datasets (e.g., across institutions, clinical protocols, and patient demographics), which need to be addressed to claim clinical impact of the models. Finally, with enough evidence collected, the model will be gradually introduced into prospective settings (i.e., clinical trials) to further validate clinical relevance, safety, and fairness, which will motivate its ultimate deployment to assist in clinical practice (see Section 7.4).

## *6.2. Data for model validation in the clinical setting*

Clinically-relevant forecasting of tumor growth and treatment response may require a diverse array of data types that can be collected throughout patient care. These data usually provide insight into the overall patient status (e.g., quality of life, cognitive status, and treatment-related side effects), tumor biology (e.g., tumor subtype, proliferation rates, and migration rates), and treatment response (e.g., overall survival and progression-free survival). As in the preclinical setting, the selection of data types largely depends on the type of model and quantities of interest under investigation (see Sections 2 and 5.2). Overall patient status can often be assessed through subjective clinical examinations and patient-reported outcomes, which provide valuable insight into otherwise intangible experiences during care (Balitsky et al., 2024; Bhatt et al., 2023). These approaches also promote a patient-centered focus, emphasizing the individual's perspective in their treatment. Insights into tumor biology can come from a wide range of data sources including laboratory-based exams and medical imaging data (Kazerouni et al., 2020). Laboratory-based exams, such as biopsies, genomic profiling, and urine or blood-based diagnostics, can provide detailed molecular, cellular, and pathological information (Basik et al., 2013; Chakravarty and Solit, 2021; Gilson et al., 2022; Hanash et al., 2011; Jordaens et al., 2023). While biopsies offer direct insights into tumor tissue, they are invasive procedures that carry risks and are often limited by sampling biases, as a single sample may not represent the heterogeneity of the tumor spatially or temporally (Basik et al., 2013; Bjurlin et al., 2014; Gilson et al., 2022; Ragel et al., 2015; Rakha and Ellis, 2007). Urine and blood-based diagnostics (Hanash et al., 2011; Jordaens et al., 2023), including liquid biopsies and biomarker assessments, provide a less invasive alternative that enables longitudinal monitoring of tumor dynamics through (for example) established clinical biomarkers (e.g., PSA and M-protein) (Cornford et al., 2024; Cowan et al., 2022) and circulating tumor cells and DNA (Alix-Panabières et al., 2012; Cescon et al., 2020). MRI, PET, and CT offer a less-invasive means of assessing tumor biology spatially and temporally (Kazerouni et al., 2020; Lorenzo et al., 2022b). Specifically, several imaging modalities and techniques have been developed to assess key tumor properties such as cell density (Padhani et al., 2011), perfusion (Quarles et al., 2019; Yankeelov and Gore, 2007), oxygenation (Perez et al., 2023), metabolism (Hofman and Hicks, 2016), and protein expression (Gao et al., 2024; Roberts et al., 2023). Importantly, imaging methods preserve the spatial context of the tumor biology enabling spatially-resolved modeling of the interaction of the tumor with different components of the patient's anatomy (e.g., healthy appearing tissue and vasculature) (Chen et al., 2012; Hormuth et al., 2021a; Lipková et al., 2019; Lorenzo et al., 2024a, 2022b, 2019a; Wong et al., 2016; Wu et al., 2022b). Insights into treatment response can come from any of the previously discussed methods, especially throughout changes with respect to baseline (pre-treatment) imaging data. Finally, large-scale databases like the Surveillance, Epidemiology, and End Results (SEER), The Cancer Imaging Archive (TCIA), and Cancer Research Data Commons provide individual and population-level data on clinical manifestation, monitoring, treatment outcomes, and survival across diverse patient groups (Clark et al., 2013; Duggan et al., 2016; Fedorov et al., 2021; Wang et al., 2024). These population data could be useful in developing and validating virtual patient cohorts (Craig et al., 2023) for evaluating optimized clinical interventions (e.g., monitoring plans, triaging criteria, and treatment regimens).

### *6.3. Examples of model validation studies in the clinical setting*

#### *6.3.1 Brain cancer*

Arguably, brain tumor growth and therapeutic response is one of the most established clinical applications of tumor forecasting in the clinical setting (Hormuth II et al., 2022; Jackson et al., 2015; Mang et al., 2020), with numerous modeling approaches informed by anatomical and/or quantitative imaging data (Agosti et al., 2018; Ayala-Hernández et al., 2021; Corwin et al., 2013; Hormuth et al., 2021a; Lipková et al., 2019; Rockne et al., 2010; Urcun et al., 2023; Zhang et al., 2025). For example, the authors in (Hormuth et al., 2021a) employed longitudinal diffusion-weighted MRI estimates of cell density to personalize a family of mathematical models describing the growth and response to chemoradiation of high-grade glioma. Through model selection using the AIC, the model that best balanced model complexity and model-data agreement was a two-species formulation describing the contrast-enhancing and non-enhancing components of the tumor. Using this model the authors further obtained less than < -2.5 % error in predicting the enhancing tumor burden and a moderate agreement at the voxel level in cell count ($CCC > 0.68$) at 3 months post-chemoradiation.

*6.3.2 Breast cancer*

Several studies have achieved validation of breast cancer models informed by imaging, histological, and/or therapeutic data to predict clinical outcomes in the setting of breast-conserving surgery, neoadjuvant chemotherapy, malignant progression, and metastatic relapse (Bigarré et al., 2023; Bosque et al., 2023; Harbin et al., 2023; Jarrett et al., 2020; Vavourakis et al., 2016; Wu et al., 2022b). In particular, the work in (Wu et al., 2022b) integrated longitudinal anatomic, diffusion-weighted, and dynamic contrast-enhanced MRI data within a mathematical model to accurately predict triple-negative breast cancer (TNBC) response to neoadjuvant chemotherapy on a patient-specific basis (see Figure 5). This model was internally validated in the dataset from the ARTEMIS trial (Yam et al., 2021) by achieving an area under the receiver operator characteristic curve (AUC, 95% confidence interval) of 0.82 (0.73 – 0.88) for differentiating pathological complete response (pCR, i.e., no residual tumor) from non-pCR TNBC patients. Furthermore, the robustness of this model was validated on the dataset from a multi-institutional trial, I-SPY2 (Barker et al., 2009), where the model achieved an AUC of 0.78 for differentiating pCR from non-pCR in a cohort of patients with various breast cancer subtypes (Patel et al., 2024).

*6.3.3 Prostate cancer*

Forecasting of prostate cancer growth and response to surgical resection, radiotherapy, hormonal therapy, and chemotherapy has also achieved various degrees of validation in the clinical setting using primarily longitudinal serum PSA data, but also imaging and histological data (Brady-Nicholls et al., 2021, 2020; Hirata et al., 2010; Lorenzo et al., 2024a, 2022a, 2019b; Morken et al., 2014, 2014; Phan et al., 2020; Sosa-Marrero et al., 2021; Vollmer, 2010; Vollmer and Humphrey, 2003; West et al., 2019; Yamamoto et al., 2016). For instance, in (Brady-Nicholls et al., 2020) the authors validated a model for the patient-specific prediction of the response of biochemically-recurrent prostate cancer to androgen deprivation therapy. The model was personalized using longitudinal PSA data and achieved a global $R^2$ of 0.63. The authors further investigated the model-informed design of optimal treatment plans with on- and off-treatment periods matching the evolutionary dynamics of the patient's tumor. These personalized treatment plans were estimated to yield a significant increase in time to progression to hormone-resistant disease *in silico*, along with an average reduction of the cumulative dose up to nearly 64% with respect to standard intermittent androgen deprivation therapy.

# 7. Open challenges in the validation of tumor forecasting models

## *7.1 Mathematical challenges*

The first challenge when using tumor forecasting to address a scientific or clinical question is to select the type and constituents of the underlying biology-informed mechanistic model. The type of formulation (i.e., ODE, PDE, or ABM) is usually decided based on the scale of the tumor-specific features available in the data and to be captured by the model (e.g., ABMs for cellular interactions, ODEs for scalar biomarker dynamics, and PDEs for three-dimensional tumor forecasts) (Brady and Enderling, 2019; Kazerouni et al., 2020; Lorenzo et al., 2023; Metzcar et al., 2019). Then, the model variables and mechanisms can be chosen based on the same two criteria, while considering previous biomedical evidence and mathematical efforts for similar scenarios (Ayala-Hernández et al., 2021; Brady-Nicholls et al., 2020; Jarrett et al., 2020; Lorenzo et al., 2024a; Strobl et al., 2022; Vega et al., 2020; Wong et al., 2016; Yang et al., 2022). An alternative approach consists of starting with the most parsimonious model and progressively incorporating mechanisms to capture more of the underlying tumor biology. For example, this strategy has resulted in successful model developments for breast (Atuegwu et al., 2013; Jarrett et al., 2020; Weis et al., 2015; Wu et al., 2022b) and brain cancer (Hormuth et al., 2021a, 2019a, 2017, 2015; Rockne et al., 2015; Swanson et al., 2011; Wang et al., 2009). Additionally, sensitivity analysis, parameter identifiability, and model selection methods help identify an adequate model formulation for the available data and the intended application (Craig et al., 2023; Hormuth et al., 2021a; Johnson et al., 2019; Lorenzo et al., 2024b, 2022b; Phillips et al., 2023b; Resende et al., 2022). Indeed, some mathematical modeling frameworks, such as OPAL (Lima et al., 2017; Lorenzo et al., 2022b; Oden et al., 2017), have integrated these methods to select the best model configuration for specific data and forecasting goals. In general, these integrated approaches first define a "parent" model including all eligible cancer mechanisms suggested by the data and intended application, then generate "daughter" models including a subset of those mechanisms, assess sensitivity and identifiability to define a minimum parameter set to be calibrated, fit each daughter model to a training dataset, validate them in a separate dataset, and finally use model selection to identify the model formulation best balancing predictive performance and complexity (e.g., number of free parameters).

## *7.2 Computational challenges*

Model validation using time series of data requiring model reparameterization, uncertainty quantification, and model-driven optimization to support clinical decision-making under uncertainty (e.g., how to treat and when to measure the tumor burden) are fundamental tasks in tumor forecasting that can require many queries of the underlying mechanistic model and lead to prohibitive computational costs. However, the timely administration of clinical interventions requires the execution of tumor forecasts and associated decision-making in clinically relevant timescales. Despite the promising growth of computational capabilities (Dongarra and Keyes, 2024), leveraging the full computational model for many-query settings such as uncertainty quantification and propagation remains intractable in many cases. This has spurred the development of algorithmic and methodological improvements to more efficiently characterize uncertainty through efficient sampling (Cotter et al., 2013) or approximation (Bui-Thanh et al., 2013; Isaac et al., 2015; Liang et al., 2023) to reduce the requisite number of model evaluations. An alternative approach which directly addresses computational cost is the development of computationally inexpensive surrogate models that approximate the high-fidelity mechanistic models. For safety-critical applications such as those in healthcare, interpretability and explainability are paramount and the surrogate model should integrate biology- and physics-based knowledge. To this end, one such approach is to fuse the predictive power of machine learning with known governing equations to develop scientific machine learning (SciML) techniques (Baker et al., 2019; Stowers et al., 2025). For example, reduced-order models (Benner et al., 2015) can encode known biology while reducing the dimensionality of the original system of differential equations, thereby maintaining the interpretability of the original mechanistic models at significantly reduced computational costs (Agosti et al., 2020; Christenson et al., 2024; Viguerie et al., 2022). Data-driven reduced-order models, such as operator inference (McQuarrie et al., 2024; Peherstorfer and Willcox, 2016; Qian et al., 2020), offer the potential of biology-aware surrogates that blend model reduction with the non-intrusive nature of data-driven learning. Other explainable surrogate modeling techniques include Gaussian process surrogates (Williams and Rasmussen, 2006) and regression (or classification) trees (Breiman, 2017). A complementary approach are multifidelity frameworks which trade-off differing levels of model fidelity and computational demand to achieve significant computational cost reduction while maintaining acceptable levels of accuracy (Peherstorfer et al., 2018). Multifidelity methods can fuse information from cheaper, lower-fidelity approximations of the original model, such as coarser discretizations, simplified biology, reduced-order models, and data-driven surrogates, to achieve model predictions supporting decision-making in clinically actionable time. In general, the appropriate approach to tackle the computational challenges depends on the tumor forecasting goals and identifying fit-for-purpose models and methods (National Academies of Sciences, 2023).

### *7.3 Data challenges*

While the abundance of biomedical data nowadays may suggest that we are 'swimming in data', existing dataset are multi-modal by nature, incomplete, non-standardized, and often misaligned with the questions to be addressed by a tumor forecasting model in development (Sweeney et al., 2023). For example, retrospective or publicly available datasets can serve as important resources model validation on an external dataset (Clark et al., 2013; Duggan et al., 2016; Fedorov et al., 2021; Wang et al., 2024).

Unfortunately, these may also be fraught with factors that can hinder their performance in model development and validation, including older or variable data standards or protocols that vary from the current standard of care, measurement noise, and diverse experimental protocols that limit their practical utility for model validation. Thus, to ensure that the data is appropriate for model validation and establishing experimental or clinical utility, it is important for effective collaboration between modelers and clinical trialists and practitioners (or preclinical researchers) to align data collection with the requirements needed for meaningful model validation (Brady and Enderling, 2019; Enderling et al., 2019; Hernandez-Boussard et al., 2021; Kazerouni et al., 2020; Lorenzo et al., 2023). This partnership must include open communication about which data or quantities of interest are most valuable to the clinical question of interest, which data types are practically accessible (e.g., cost, invasive vs. non-invasive, and availability), and the timing and frequency of the data collection. Nevertheless, clinical settings often face significant constraints regarding data availability as many critical variables cannot be measured as frequently or comprehensively as desired by modelers, particularly in vulnerable or acutely ill populations. To address this issue, approaches such as optimal experimental design (Cho et al., 2020; Gevertz and Kareva, 2024) could guide the selection of data types, timing, and frequency to maximize the information gained from them in the modeling framework.

## *7.4 Clinical challenges*

If the technical hurdles described in Sections 7.1 – 7.3 can be addressed, then it is possible to develop a predictive, mathematical model built on relevant biological mechanisms implemented with a reliable and efficient numerical implementation that is capable of being personalized by clinically available data. Assuming that such a model has been successfully validated with internal and external data, the final stage of model validation is the (likely) very challenging step to deploy the model to help guide clinical decision-making in daily practice. More specifically, it would be necessary to design appropriate clinical trials to validate the benefits of incorporating the predictive models into the clinical decision-making process. It is important to stress that both the concept of tailoring treatment on a patient-specific basis and the quantitative tools to guide the intervention represent novel departures from conventional clinical trial design (Hariton and Locascio, 2018; Yankeelov et al., 2024). Making this paradigm change will require building trust among clinicians and patients, collecting more retrospective evidence of the predictive power of tumor forecasts, and performing prospective trials designed to directly compare the benefits of personalized tumor forecasting with respect to the standard of care at the population level (i.e., whether the patient-specific predictions significantly improve the clinical management of the disease across the diverse patient population for the clinical scenario of interest). Fortunately, there is an increasing interest in the notion of model-guided cancer therapy. For example, the authors in (Lin et al., 2023) developed a software-aided imaging technique leveraging a biomechanical deformable registration model (Morfeus) to intra-procedurally support confirmation of tumor coverage during ablation therapy of liver cancer. Retrospective validation demonstrating this technique has shown that the accuracy in quantifying minimal ablative margin (Lin et al., 2024) and predicting local tumor progression can be significantly improved (Lin et al., 2023). A phase II trial (NCT04083378 (Lin et al., 2022)) was also performed to prospectively assess the clinical benefits of the model, including improving local tumor progression-free survival, overall survival, and quality of life (Lin et al., 2024). The success of this study is built on sufficient retrospective evidence and close collaboration between modelers and clinicians from the earliest stages of designing the clinical trial. Although the specific model-guided decision is personalized, this clinical trial followed a conventional randomized two-arm design with the use of the personalized technology as the experimental arm, thereby allowing for statistical evaluation of effectiveness. Further developments in novel adaptive trial diagrams (Bhatt and Mehta, 2016; Pallmann et al., 2018) and statistical metrics to analyze individualized treatment effects (Imai and Li, 2023; Künzel et al., 2019) also provide opportunities to address challenges in the clinical validation of personalized predictive models.

## 8. Conclusion

The science of tumor forecasting has experienced intense development over the last decade. Mathematical models of cancer growth and treatment response have been shown to predict patient-specific disease outcomes across different tumor sites and clinical scenarios (e.g., disease monitoring and response to radiotherapy or chemotherapy), with some studies further proposing the use of tumor forecasts to optimize clinical management of the disease for individual patients. Additionally, mathematical models of cancer have enabled systematic exploration of the biophysical phenomena underlying tumor biology in a plethora of preclinical scenarios, and their predictions have further been proposed to plan optimal experimental studies that interrogate specific mechanisms underlying cancer growth and treatment response. Hence, the advance of cancer research and clinical oncology from the current observational, population-based standard toward a predictive, individualized paradigm is inevitable. In this context, model validation is a fundamental requirement to establish trust in the tumor predictions. This goal is of utmost importance for the high-stakes decisions present in clinical oncology, as well as for fruitfully guiding preclinical research efforts to unveil new facets of cancer development and treatment response. Thus, research efforts in the field need to account for adequate models and validation metrics, efficient computational implementations that are fit-for-purpose and enable rapid decision-making in clinically-relevant times, as well as quantification of the inherent uncertainty in tumor measurements, model predictions, and the decision made based on them. Ultimately, cancer modelers will further need to devise plausible strategies to achieve ultimate validation of tumor forecasting in clinical trials and, hence, bring personalized predictions from the computer to the bedside.

# Figures

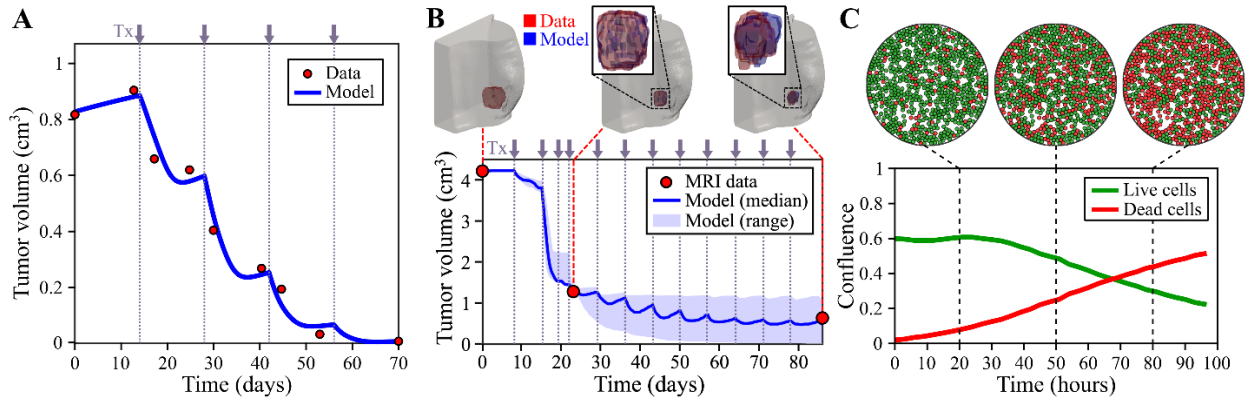

**Figure 1. Examples of main outputs from ODE models, PDE models, and ABMs of tumor growth and treatment response.** Panel A shows a characteristic prediction of an ODE model describing the response of a tumor to cytotoxic chemotherapy (Tx; i.e., a treatment with a drug that directly induces tumor cell death) in terms of the temporal dynamics of tumor volume (Lorenzo et al., 2024b, 2023). During each treatment cycle, the tumor volume reduces in size due to the cytotoxic effect of the drug. However, once most of the drug has been removed from the patient's body, the tumor resumes growth before the administration of the next cycle. Panel B illustrates a prediction of a PDE-based model of breast cancer response to neoadjuvant chemotherapy (Tx). The model is initialized and calibrated within the patient's breast anatomy using longitudinal MRI (Wu et al., 2022b). Spatial integration of the region occupied by the tumor yields the tumor volume. This scalar quantity of interest can be calculated over time from the spatiotemporal model predictions, resulting in a similar plot as in panel A. Furthermore, this example accounts for uncertainty, which is represented as a range around the median of the model predictions. Panel C illustrates the output of an ABM that describes the dynamics of a tumor cell colony in a microenvironment with low oxygen concentration. The latter, combined with the initial high number of cells, is insufficient to sustain the growth of the tumor cell population and, hence, induces tumor cell death over the course of the simulation. The calculation of confluence (i.e., the fraction of the well occupied by each type of tumor cell) provides a time-resolved scalar quantity of interest (similarly to that panel A) that further enables comparison with standard experimental results (Lima et al., 2021).

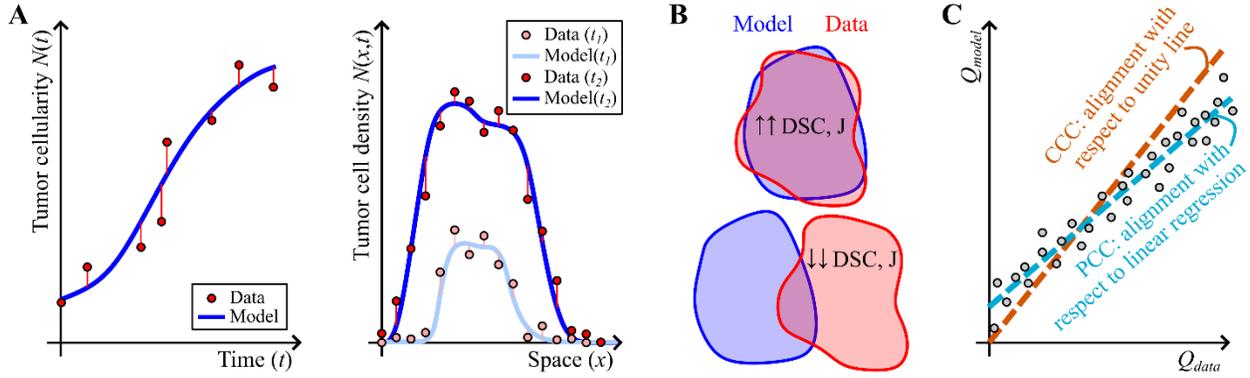

**Figure 2. Example of common metrics used for model validation.** Panel A illustrates how error metrics, such as RMSE, MAE, and MAPE, measure the agreement between model predictions and data. For time-resolved quantities of interest (left plot), these metrics calculate a distance between the values produced by the model (blue curve) and the datapoints (red bullet points). These distances are then summed to produce an overall metric of model fitting or forecasting error. For spatiotemporally-resolved quantities of interest (right plot), the same approach can be applied over the spatial dimensions at a specific timepoint or it can be further extended across all timepoints of available data. Panel B illustrates how DSC and J capture geometric misalignments in the shape of the tumor measured from spatiotemporal data (e.g., imaging) and from a spatiotemporal model (e.g., PDEs and ABM). The closer the two shapes obtained from data and model forecasts, the higher the DSC and J values will be. Panel C presents the information provided by the correlation coefficients PCC and CCC. In the unity plot depicted in this panel, each point is determined by the value of a quantity of interest calculated from the model ($Q_{model}$) and measured from data ($Q_{data}$). While the PCC measures to what extent these points exhibit a positive linear trend, the CCC specifically measures alignment with respect to the unity line (i.e., $Q_{model} = Q_{data}$). As explained in Section 4.3, the metrics depicted in this figure can also be leveraged for uncertainty quantification of the model predictions and, thus, for probabilistic model validation.

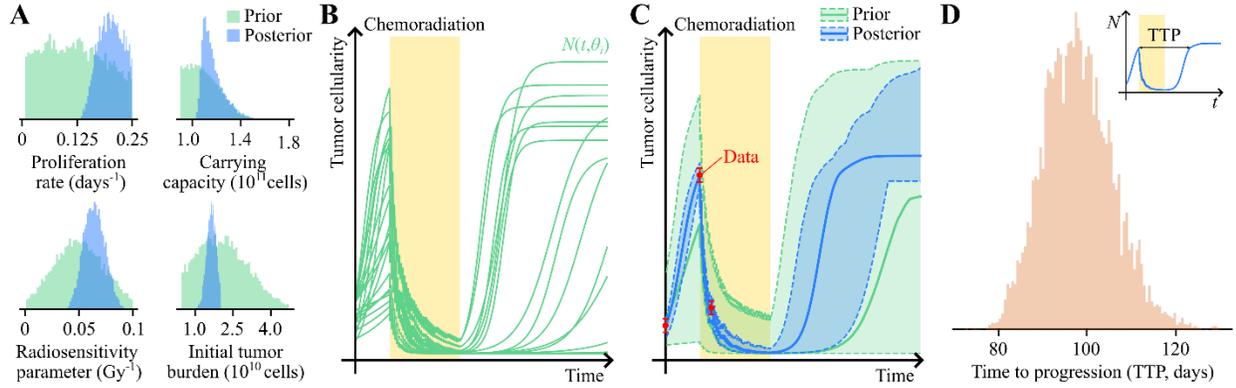

**Figure 3. Uncertainty quantification.** Panel A illustrates the prior distributions ($\pi_{prior}(\theta)$) and patient-specific posterior distributions ($\pi_{post}(\theta|d)$) of the parameters of an ODE model of high-grade glioma response to chemoradiation (Chaudhuri et al., 2023). The prior distributions were built from data in the literature, while the patient-specific posterior distributions were calculated using longitudinal patient measurements of tumor cellularity *via* Bayesian calibration (see Eq. [15]). Panel B exemplifies the different dynamics of tumor cellularity that can be obtained with the ODE model by sampling different parameter values from the prior distribution. Panel C shows the patient-specific posterior distribution of tumor cellularity forecasts over time obtained from sampling the parameters of the ODE model from the patient's posterior distributions as well as from the prior distributions. The solid lines represent the median forecast, while the dashed lines represent the minimum and maximum. The predictive posterior distribution of tumor cellularity values can be compared against the data (and their error distribution) at the times of available measurements to establish probabilistic validation of the model. Panel D shows the distribution of the time to propagation (TTP), which is a quantity of interest calculated from the distribution of forecasts as the time elapsed between treatment onset and the time at which the tumor is predicted to regrow to the same tumor cellularity existing at the beginning of treatment. The analysis of this TTP distribution under alternative treatment plans for each individual patient can provide guidance for optimal therapeutic decision-making under uncertainty (Chaudhuri et al., 2023).

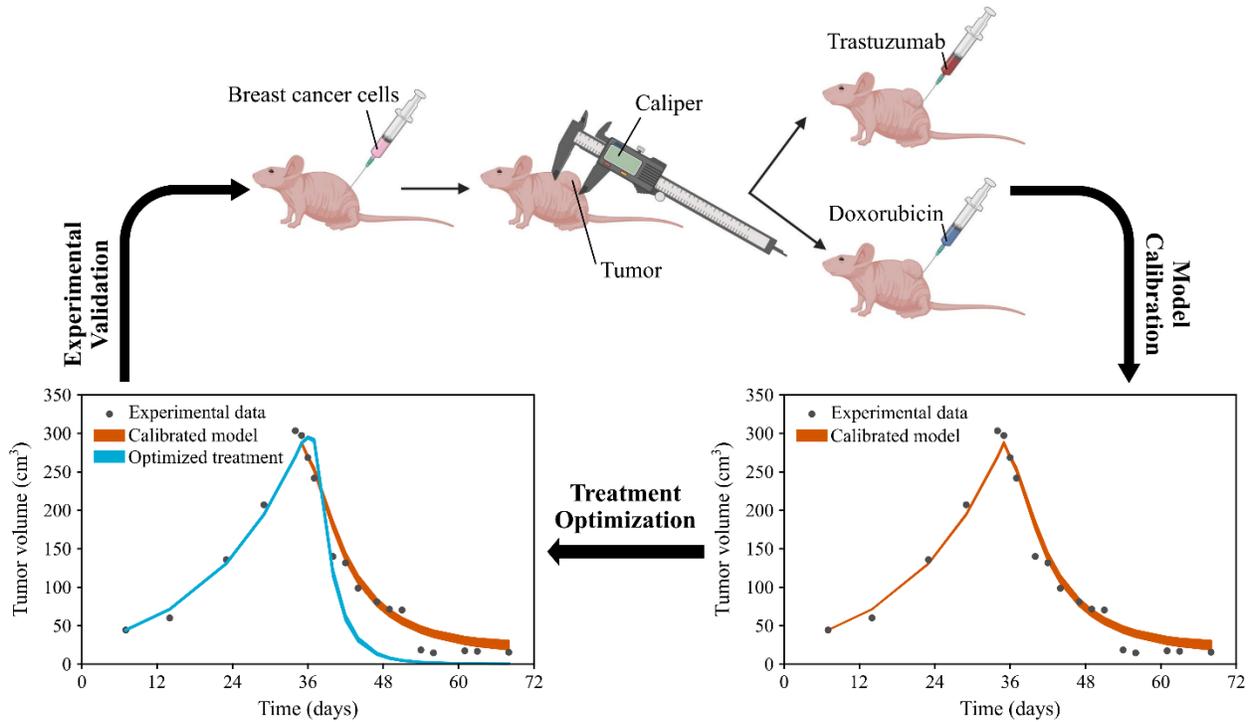

**Figure 4. Example of an experimental setup for model validation and model-informed treatment optimization.** The preclinical study shown in this figure aims at finding optimal treatment plans for HER2+ breast cancer consisting of combinations of trastuzumab (a standard-of-care drug targeting HER2+ breast cancer) and doxorubicin (a standard-of-care cytotoxic drug). To this end, a mathematical model of breast cancer growth and response to this combined treatment is developed and validated using experimental measurements of tumor volume in mice that were injected with HER2+ breast cancer cells. These measurements are regularly collected using a caliper before and after treatment, as shown by the data points exhibiting a rising and declining trend in the bottom plots, respectively. Then, the validated model (orange curve) is further leveraged to design an optimal treatment protocol that maximizes tumor control (i.e., minimum tumor volume during the treatment phase of the experiment) while maintaining the same total drug dose as in the experiment (blue curve). Finally, the optimal protocol is validated in a new set of experiments. This methodology and setup are based on the study in (Lima et al., 2022), where further methodological details are available. This figure was partially created using BioRender.com.

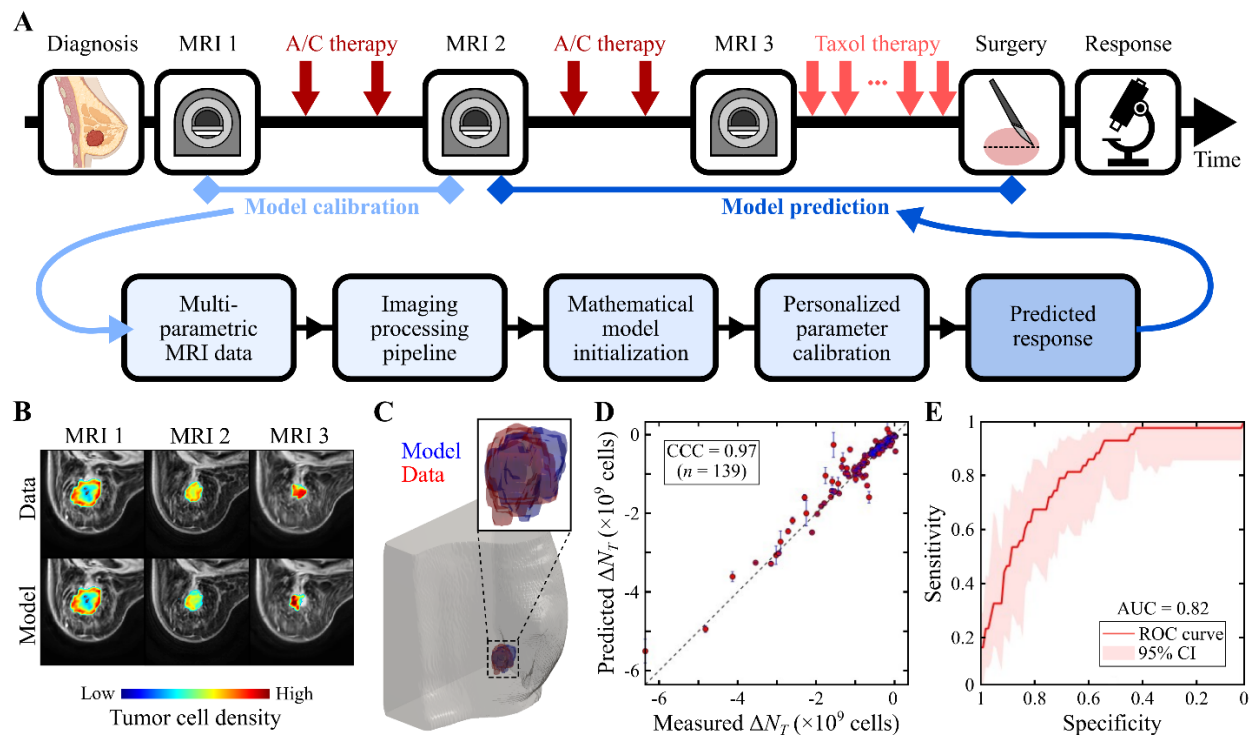

**Figure 5. Example of a clinical scenario for model validation.** The goal of the tumor forecasting study shown in this figure is to predict the response of TNBC to neoadjuvant chemotherapy. The treatment usually consists of four cycles of Adriamycin® and Cytoxan® (A/C; i.e., doxorubicin and cyclophosphamide) every two weeks and twelve weekly doses of Taxol® (i.e., paclitaxel), with the order of these two drug regimens being exchangeable in clinical practice (Patel et al., 2024; Wu et al., 2022b). Panel A illustrates the clinical setup and the integration of patient data within the model. In this example, following diagnosis, the patient had three multiparametric MRI scans: before treatment onset, between the second and third cycles of A/C, and after the conclusion of A/C therapy. Following termination of Taxol® therapy, the patient underwent surgery and the excised tissue was evaluated by histopathology to determine whether the patient achieved pathological complete response (pCR; i.e., no residual tumor) or not (i.e., non-pCR). Predicting this outcome early in the course of treatment would enable the treating physician to adjust treatment to maximize tumor burden reduction while controlling for minimal toxicity. To this end, a PDE-based mathematical model is initialized and calibrated using multiparametric MRI data from the first and second imaging visits. The resulting personalized model is leveraged to make a prediction from the time of the second MRI to the time of surgery. Panel B shows an example of the patient-specific tumor cell density maps measured from MRI and predicted by the model at the three imaging visits. In this case, the patient had Taxol® therapy first and the second MRI was acquired between the third and fourth cycles. Panel C depicts a 3D visualization of the patient-specific breast geometry and tumor at the conclusion of Taxol® chemotherapy for the same patient as in panel B. Panel D provides a plot of the correlation between the measured and predicted changes
32

in total tumor cellularity ($N_T$) up to the end of A/C chemotherapy for a group of TNBC patients who received A/C as their first drug reimen. Panel E depicts the receiver operator characteristic curve analysis of differentiating pCR from non-pCR based on predicted residual tumor burden at the end of the entire neoadjuvant chemotherapy regimen for the same patient cohort as in panel D. This methodology and setup are based on the studies in (Patel et al., 2024; Wu et al., 2022b), where further methodological detail is available. This figure was partially created using BioRender.com.

## Acknowledgments

GL acknowledges grant PID2023-146347OA-I00 funded by MICIU/AEI/10.13039/501100011033 and ERDF/EU, as well as grant RYC2022-036010-I funded by MICIU/AEI/10.13039/501100011033 and ESF+. DH acknowledges support from CPRIT 220225 and NSF DMS 2436499. AC, GP, and KW acknowledge funding from DOE ASCR award DE-SC002317, NSF FDT-Biotech award 2436499, and DARPA award DE-AC05-76RL01830. This material is based upon work supported by the National Science Foundation Graduate Research Fellowship under Grant No. DGE2137420 (RJSP). TEY acknowledges the National Cancer Institute for funding through 1U24 CA226110, 1R01CA240589, 1U01CA253540, 1R01CA260003, 1R01CA276540 and NSF 2436499.